\documentclass[aps,pre,preprint,superscriptaddress,showpacs]{revtex4}
\usepackage{graphicx,epstopdf}
\begin{document}

% Use the \preprint command to place your local institutional report
% number in the upper righthand corner of the title page in preprint mode.
% Multiple \preprint commands are allowed.
% Use the 'preprintnumbers' class option to override journal defaults
% to display numbers if necessary
\preprint{LA-UR-07-8103}

%Title of paper
\title{Kinetic Monte Carlo Method for Rule-based Modeling of Biochemical Networks}

% repeat the \author .. \affiliation  etc. as needed
% \email, \thanks, \homepage, \altaffiliation all apply to the current
% author. Explanatory text should go in the []'s, actual e-mail
% address or url should go in the {}'s for \email and \homepage.
% Please use the appropriate macro foreach each type of information

\author{Jin Yang}
\email[Email Address: ]{yangjin@picb.ac.cn}
\affiliation{CAS-MPG Partner Institute for Computational Biology, Shanghai Institutes for Biological Sciences, Chinese Academy of Sciences,  Shanghai 200031, China.}
\author{Michael I. Monine}
\affiliation{Theoretical Division and Center for Nonlinear Studies, Los Alamos National Laboratory, Los Alamos, NM 87545, USA.}
\author{James R. Faeder}
\email[Email Address: ]{faeder@pitt.edu}
\affiliation{Department of Computational Biology, University of Pittsburgh School of Medicine, Pittsburgh, PA 15260, USA.}
\author{William S. Hlavacek}
\email[Email Address: ]{wish@lanl.gov}
\altaffiliation{Department of Biology, University of New Mexico, Albuquerque, NM 87131, USA.}
\affiliation{Theoretical Division and Center for Nonlinear Studies, Los Alamos National Laboratory, Los Alamos, NM 87545, USA.}

\date{Version 15: \today}
\begin{abstract}
We present a kinetic Monte Carlo method for simulating chemical transformations specified by reaction rules, which can be viewed as generators of chemical reactions, or equivalently, definitions of reaction classes.  A rule identifies the molecular components involved in a transformation, how these components change, conditions that affect whether a transformation occurs, and a rate law.  The computational cost of the method, unlike conventional simulation approaches, is independent of the number of possible reactions, which need not be specified in advance or explicitly generated in a simulation.  To demonstrate the method, we apply it to study the kinetics of multivalent ligand-receptor interactions.  We expect the method will be useful for studying cellular signaling systems and other physical systems involving aggregation phenomena.
\end{abstract}

\pacs{82.39.Rt,87.15.R-,87.17.Aa,87.16.Xa,02.70.Tt,05.10.Ln}
\maketitle

%%%% BEGIN INTRODUCTION

Proteins in cellular regulatory systems, because of their multicomponent composition, can interact in a combinatorial number of ways to generate myriad protein complexes, which are highly dynamic \cite{Hl03}.  This feature of protein-protein interactions has been called combinatorial complexity, and it is recognized as a major barrier to understanding cell biology \cite{Hl03,En01,Br03,Kh06}.  The problem of combinatorial complexity is alleviated by using a rule-based approach to model protein-protein interactions \cite{Hl06}.  In this approach, proteins and protein complexes are represented as structured objects (graphs) and protein-protein interactions are represented as (graph-rewriting) rules that operate on these objects to modify their properties, consistent with transformations mediated by the interactions being represented.  Rules can serve as definitions of individual reactions or entire reaction classes, and they can be used as generators of reactions \cite{Bl04,Fa05a}.  The assumption underlying this modeling approach, which is consistent with the modularity of regulatory proteins \cite{Pa03}, is that interactions are governed, at least to a first approximation, by local context that can be captured in simple rules (e.g., by the availability of binding sites on two binding partners).  Rules can, in principle, be used to generate reaction networks that account comprehensively for the consequences of specified protein-protein interactions.  However, the size of a rule-derived network can severely challenge conventional methods for simulating reaction kinetics \cite{Hl06}.  For example, the rule set formulated by Danos et al. \cite{Da07a} implies more than $10^{23}$ chemical species and an even greater number of reactions.  

It is impractical to simulate the kinetics of such a rule-derived network with the methods that are most commonly used in modeling studies of cellular regulatory systems, such as Gillespie's method \cite{Gi76,Gi77}.  These methods tend to be ones that are applicable in the well-mixed limit, and they are generally population based, meaning that they explicitly track populations of chemical species.  The computational cost of simulation is $O(\log_2 M)$ per reaction event for efficient kinetic Monte Carlo (KMC) implementations \cite{Vo07,Gi07}, where $M$ is the number of  reactions.  For integration of ordinary differential equations (ODEs) derived from the law of mass action, the cost is polynomial in the number of chemical species and typically cubic for stiff ODEs.  In addition to the cost of simulation, the cost of generating a network from rules, which is necessarily incurred either before or during simulation \cite{Fa05a,Lo05,Gi07}, can be prohibitively expensive.  One reason for the expense of network generation is that the product(s) of a new reaction derived from a rule must be compared with the chemical species stored in computer memory to establish uniqueness, which requires graph isomorphism checking if one uses graphs to track the connectivity of proteins \cite{Bl06}.  Another barrier to simulation is simply the amount of memory required to store the chemical species and reactions that form a large-scale network.  

To address these computational limitations, Krivine, Danos and co-workers \cite{Da07b} have developed a particle-based method that is suitable for simulating the kinetics of cellular regulatory systems and other systems for which chemical transformations can be defined in terms of reaction rules.  This method, which we will refer to as the DFFK method, avoids the expense of network generation by directly using rules to propagate a stochastic, discrete-event simulation in which molecules undergo transformations sampled from rule-defined reaction classes.  The cost of the DFFK method is a function of $m$, the number of rules, rather than $M$, the number of reactions that can be generated by the rules.  Memory requirements are also independent of $M$.  For $m \ll M$, the computational cost of tracking the states of individual molecules can be far less than that associated with tracking the chemical species that these molecules (potentially) populate.  The DFFK method is closely related to various other simulation methods that have been developed mainly for application to non-biological systems \cite{Fr95,Sch04,Ja05,Ch07,Sch08,Sl08}.  For example, Schulze \cite{Sch04,Sch08} has described a method for stochastic simulation of crystal growth that is applicable when the number of distinct reaction rates in a system is less than the number of reactions, which is exactly the scenario considered in a rule-based description of protein-protein interaction kinetics.  Another notable method is that of Slepoy et al. \cite{Sl08}.  Both of these methods have a computational cost that is independent of $M$.

Here, we present an extension of the DFFK method, which we call the rule-based KMC method.  The method allows for imposition of contextual constraints specified in a rule on the rates of reactions defined by the rule.  In other words, the rate associated with a transformation defined by a rule can be adjusted to account for the molecular context of the transformation.  This capability is important for modeling aggregation, as will be seen below, and other phenomena \cite{Ba07}.

To demonstrate the rule-based KMC method, we apply it to simulate a rule-based model that characterizes the interaction kinetics of a population of trivalent ligands with a population of bivalent cell-surface receptors (Fig.~1).  This model, which we will call the TLBR model, is relevant for studying a number of experimental systems that have recently been reported in the literature \cite{Po02,Bi07,Po07,Si07}. We have formulated the TLBR model, a kinetic model, so that it corrresponds to the equilibrium model of Goldstein and Perelson \cite{Go84}, which can be used to characterize the equilibrium behavior of the TLBR model in the continuum limit.  The equilibrium model predicts a sol-gel region, in which a macroscopic fraction of the receptors are found in a single giant aggregate.  As the percolation transition is approached, and the mean size of ligand-induced receptor aggregates increases, the number of distinct reactions that can occur explodes, which prohibits simulation of the reaction kinetics using population-based methods near or in the sol-gel region.  Simulation of the TLBR model is a challenging and ideal test case for the rule-based KMC method, because the number of reactions that have a non-zero stationary flux can be tuned over a broad range by adjusting the model parameters that control mean aggregate size, which is limited only by total receptor number.  Moreover, to obtain correct simulation results, one requires the extension of the DFFK method that is presented here.

%%%%%FIGURE 1
\begin{figure}
\centering
\includegraphics[width=3.2in]{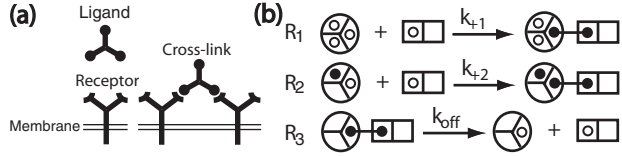}
\caption{\label{fig:TLBR} TLBR model. (a) A ligand with three identical binding sites and a mobile cell-surface receptor with two identical binding sites.  The ligand mediates cross-linking of receptors as shown.  (b) Rules representing capture of a freely diffusing ligand by a receptor ($R_1$), ligand-mediated receptor cross-linking ($R_2$), and ligand-receptor dissociation ($R_3$).  Parameters of the rate laws associated with these rules are single-site rate constants: $k_{+1}$, $k_{+2}$, and $k_{\text{off}}$, respectively.  An empty (filled) circle indicates a free (bound) site, a line connecting circles indicates a bond, and an empty box or wedge indicates a site that may be either free or bound.  In BNGL \cite{Fa08}, the rules are specified as follows: $R_1$ is {\tt L(r,r,r) + R(l) -> L(r!1,r,r).R(l!1)}, $R_2$ is {\tt L(r!+,r) + R(l) -> L(r!+,r!1).R(l!1)}, and $R_3$ is {\tt L(r!1).R(l!1) -> L(r) + R(l)}, where {\tt l} and {\tt r} are used to represent binding sites of the receptor ({\tt R}) and ligand ({\tt L}), respectively.}
\end{figure}

%%%% BEGIN DISCUSSION OF PHYSICAL SYSTEMS and REPRESENTATIONAL FRAMEWORK 

We consider a well-mixed reaction compartment of volume $V$ containing a set of molecules $P=\{P_1,\ldots,P_N\}$, which we take to be proteins or other molecules comprised of a set of components $C=\{C_1,\ldots,C_n\}$.  Each component $C_i$ has a local state, denoted $S_i$, that includes its type, binding partner(s), which (if any) are other components, and internal state(s), which may represent conformations or covalent post-translational modifications.  (A regulatory protein typically undergoes modifications, such as phosphorylation of a tyrosine residue, that affect its function but not its essential identity.)  The state of a protein is determined by its set of components and their states.  The state of the whole system is given by $P$, $C$, and the set of component states $S=\{S_1,\ldots,S_n\}$. 

Molecules interact according to a set of reaction rules $R=\{R_1,\ldots,R_m\}$.  Precise specification of rules is possible using established syntactic and semantic conventions, such as $\kappa$-calculus \cite{Da04}, BNGL \cite{Bl06,Fa08}, or $\rho_{\rm bio}$-calculus \cite{An07}.  Here, we adopt functional definitions that do not depend on the specific details of these conventions.  A rule $R_i$ defines necessary local and global features of $M_i$ reactants, a transformation (of molecularity $M_i$) that changes the state of $N_i$ types of components, and a rate law $r_i$ from which the {\em maximum} cumulative rate of all reactions implied by the rule can be determined.  The local features specified in a rule provide criteria for selecting components that can potentially react based on the individual properties of reactants (e.g., the states of components in a molecule), whereas the global features specified in a rule, which are optional, provide criteria for adjusting the rate at which selected components react based on the joint properties of reactants (e.g., the connectivity of two molecules).  For evaluation of rate laws, each rule $R_i$ is associated with $N_i$ sets of reactive components, denoted $X_{ij}$ for $j=1,\ldots,N_i$.  Components in $X_{ij}$ are all of the same type and each has properties consistent with local features specified in rule $R_i$.  A simple example of a rate law is that for an elementary bimolecular association reaction in which two complementary components bond ($M_i=N_i=2$): $r_i=v_i \prod_{j=1}^{M_i} |X_{ij}|$, where $|X_{ij}|$ denotes the number of components in $X_{ij}$ and $v_i$ represents the maximum rate at which a pair of components in $X_{i1} \times X_{i2}$ undergoes transformation according to $R_i$.  We note that some of the pairs in $X_{i1} \times X_{i2}$ may react at lower or even zero rate depending on the global features specified in the rule, which essentially provide rule application conditions.  As explained below, by taking advantage of the distinction between local and global features, we can sample a bimolecular or higher-order class of reactions without forming the set of combinations of reactive components.

Examples of reaction rules are illustrated in Fig.~\ref{fig:TLBR}, which presents the complete set of rules that define the TLBR model.  Rule $R_1$ is associated with two sets of reactive components: $X_{11}$, the set of ligand binding sites on free ligand molecules, and $X_{12}$, the set of free receptor sites.  Rule $R_2$ is associated with $X_{21}$, the set of free ligand binding sites on receptor-associated ligands, and $X_{22}$, which is identical to $X_{12}$.  Rule $R_3$ is associated with $X_{31}$, the set of bound ligand binding sites, and $X_{32}$, the set of bound receptor binding sites.  A bijective mapping relates the elements of $X_{31}$ and $X_{32}$.  The rate laws associated with the three rules are $r_1= (k_{+1}/V) |X_{11}| \cdot |X_{12}|$, $r_2=(k_{+2}/V) |X_{21}| \cdot |X_{22}|$, and $r_3=k_{\text{off}} |X_{31}| = k_{\text{off}} |X_{32}|$.  In $R_1$ and $R_2$, the plus sign on the left-hand side of the arrow indicates a molecularity of 2, which limits application of $R_2$ to cases where ligand and receptor binding sites are unconnected.  In other words, in the TLBR model, sites within the same ligand-receptor complex are considered to be non-reactive, which prevents the formation of cyclic aggregates, consistent with simplifying assumptions of the equilibrium version of the model \cite{Go84}.  (Extension of the TLBR model to account for cyclic aggregates, such as those suggested by the data of Whitesides and co-workers \cite{Bi07}, is beyond the intended scope of this report.)  When large aggregates form, the connectivity check needed to avoid formation of cyclic aggregates can be expensive, as we discuss below.

%%%% BEGIN DESCRIPTION OF ALGORITHM

We now describe a KMC algorithm for propagating a system $(P,C,S)$ under the influence of $R$.  Initialization requires that $(P,C,S)$ be used to construct $X$, all sets of reactive components associated with rules, and that $X$ be used to calculate the (maximum) rates given by $r$, the set of rate laws associated with rules.  In describing the method used to determine the time of the next event in a simulation and the rule to apply, we follow Gillespie's (direct) method \cite{Gi76,Gi77} for convenience of presentation with the understanding that various optimizations are possible \cite{Gi07,Li08}.  A set of rules generates events in a Poisson-distributed manner, just as a set of reactions in a conventional stochastic simulation \cite{Fi91}, and thus, essentially the same procedures can be used.  The waiting time, $\tau$, to the next event is given by
\begin{equation}
\tau = -(1/r_{\text{tot}})\ln(\rho_1)
\label{eq:tau}
\end{equation}
where $r_{\text{tot}}=\sum_{j=1}^m r_j$ and $\rho_1 \in (0,1)$ is a uniform deviate.  Next a rule $R_J$ to apply is selected by finding the smallest integer $J$ that satisfies
\begin{equation}
\sum_{j=1}^J r_j > \rho_2 r_{\text{tot}}
\label{eq:j}
\end{equation}
where $\rho_2 \in (0,1)$ is a second uniform deviate.  The cost of finding $J$ in this way is $O(m)$, so for larger values of $m$ one may wish to use a more efficient procedure that reduces the cost to $O(\log_2 m)$ \cite{Bl95,Gi00}.  Next, the particular reactants to which $R_J$ is applied are determined by selecting one component $x_k$ randomly from each set $X_{Jk}$ for $k \in \{1,\ldots,N_J\}$.  The next step extends the DFFK method.  To determine whether the selected components react, the application conditions of $R_J$ derived from the global features that it specifies are evaluated to determine an adjusted rate of reaction, $v_J^{\prime}$, which is then compared against the maximal rate of reaction, $v_J$.  If $v_J^{\prime}> \rho_3 v_J$, where $\rho_3 \in (0,1)$ is a uniform deviate, the transformation specified by the rule is applied to the selected reactants.  Otherwise, a null event occurs, i.e., a time step without a reaction.  Time is updated by setting $t \leftarrow t + \tau$ regardless of whether a reaction occurs because the sampling rate $r_{\text{tot}}$ includes non-reactive contributions.  The maximum number of random deviates that must be generated is $N_J+3$.  We now update $(P,C,S)$ and $X$ and recalculate cumulative rates $r$.  The simulation procedure outlined above is iterated until a stopping criterion is satisfied.

The above algorithm is used as follows to simulate the TLBR model.  We specify parameters: the system volume $V$, the rate constants $k_{+1}$, $k_{+2}$ and $k_{\rm off}$, and the total numbers of ligands ($N_L$) and receptors ($N_R$).  If all ligands and receptors are initially free, then all ligand sites (three per ligand) are assigned to set $X_{11}$ and all receptor sites (two per receptor) are assigned to set $X_{12}$ at time $t=0$.  All other sets associated with the rate laws of rules (e.g., $X_{21}$, $X_{31}$ and $X_{32}$) are empty.  Recall that sites and molecules are tracked individually (i.e., they are each assigned a unique label), and note that we can use $X_{12}$ in place of $X_{22}$ whenever necesssary.  The values of $r_1$, $r_2$, and $r_3$ are calculated using the expressions given earlier.  At $t=0$, $r_1=6(k_{+1}/V)N_LN_R$, $r_2=0$ and $r_3=0$. Equation~\ref{eq:tau} is used to select a time step $\tau$.  Equation~\ref{eq:j} is used to select a rule.  If $R_1$ is selected, a site $x_1$ in $X_{11}$ and a site $x_2$ in $X_{12}$ are randomly selected and reassigned to $X_{31}$ and $X_{32}$, respectively.  The mapping between $X_{31}$ and $X_{32}$ is updated to link these sites (and the molecules of which they are members) to each other.  Then, the other two sites on the ligand containing $x_1$ are assigned to $X_{21}$.  A similar process occurs if rule $R_3$ is selected.  Rules $R_1$ and $R_3$ generate no null events because pairs of sites that react according to these rules can be identified on the basis of their local features alone. In contrast, $R_2$ generates null events because pairs of sites that react according to $R_2$ must be identified on the basis of both their local and global features.  If $R_2$ is selected, a site $x_1$ in $X_{21}$ and a site $x_2$ in $X_{22}$ ($=X_{12}$) are randomly selected.  At this point, the mapping between $X_{31}$ and $X_{32}$ is used to determine if $x_1$ and $x_2$ are indirectly connected.  If not, $x_1$ is reassigned to $X_{31}$, $x_2$ is reassigned to $X_{32}$, and the mapping between $X_{31}$ and $X_{32}$ is updated to link $x_1$ and $x_2$.  If $x_1$ and $x_2$ are found to be connected, no reaction (i.e., a null event) occurs.  Finally, time is incremented.  The procedure described above is repeated, beginning with the selection of a new time step.  Execution ends when the current time exceeds a specified value. By storing the sets $X_{11}$, $X_{12}$, $X_{21}$, $X_{31}$, $X_{32}$ and the mapping between the sites of $X_{31}$ and $X_{32}$ in memory at desired time points, the kinetics of any molecular property of interest can be determined after simulation is complete.

The computational cost of the above procedure without the step of checking a rule application condition has been carefully analyzed by Danos et al. \cite{Da07b}.  The worst-case bound on cost for an efficient implementation is proportional to $\log_2 m$ plus a constant cost that is a well-defined function of certain properties of $R$, the set of rules under consideration, but not the rate laws associated with rules. In contrast, the cost of checking a rule application condition, as we will see, can depend on properties of the chemical reaction network implied by a set of rules, which in turn depend on the rate laws associated with rules.  

%%%%% DISCUSSION OF THE TLBR PROBLEM BEGINS HERE

\begin{figure}
\centering
\includegraphics[width=2.0in]{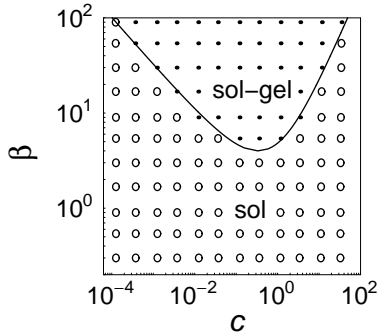}
\caption{\label{fig:PT} Percolation transition between sol and sol-gel regions in the space of $c$ and $\beta$.  The curve marks the percolation transition boundary according to the equilibrium continuum model of Goldstein and Perelson \cite{Go84}. Using the rule-based KMC method, we simulated the TLBR model to determine the steady-state value of $f_g$, the fraction of receptors in the gel phase (i.e., in the largest receptor aggregate), as a function of $c$ and $\beta$. At points marked by dots, $f_g \geq 0.05$, whereas at points marked by circles, $f_g<0.05$. To adjust the values of $c$ and $\beta$, we varied $k_{+1}$ and $k_{+2}$ and held other parameters constant at the following values: $N_R=3,000$, $N_L=42,000$, and $k_{\text{off}}=0.01$ s${}^{-1}$.}
\end{figure}

We now apply the rule-based KMC method to study the TLBR model (Fig.~\ref{fig:TLBR}).  The equilibrium receptor aggregate distribution is controlled by two dimensionless parameters: $c_{\text{tot}} = 3k_{+1}N_L/k_{\text{off}}$, or equivalently $c=3k_{+1}L_0/k_{\text{off}}$, and $\beta=k_{+2}N_R/k_{\text{off}}$ \cite{Go84}, where $L_0$ is the number of free ligands at equilibrium. The sol-gel coexistence phase predicted by the equilibrium model forms a U-shaped region in the phase diagram plotted as $\beta$ versus $c_{\text{tot}}$ (or $c$), and for a given value of $c_{\text{tot}}$ (or $c$), aggregation increases monotonically with $\beta$, and the gel (i.e., infinite cluster of receptors) appears when $\beta$ exceeds a critical value \cite{Go84}. Rule-based KMC simulations were used to recapitulate the entire phase diagram reported in Fig. 7 of \cite{Go84} (Fig.~\ref{fig:PT}).  A variety of other equilibrium properties were calculated and found to agree with the equilibrium model after accounting for the effects of finite system size (not shown).  These results confirm the validity of the rule-based KMC method.  

%%%%%FIGURE 2

To demonstrate the efficiency of rule-based KMC relative to that of population-based methods, which require reaction network specification, we will focus on one population-based method, the approach of on-the-fly simulation \cite{Fa05a,Lo05,Gi07}.  This approach is a stochastic simulation method that is designed to minimize the cost of generating a reaction network from rules.  Lazy evaluation of rules is used to generate only the part of a network that is relevant for advancing a simulation.  

\begin{figure}
\includegraphics[width=3.25in]{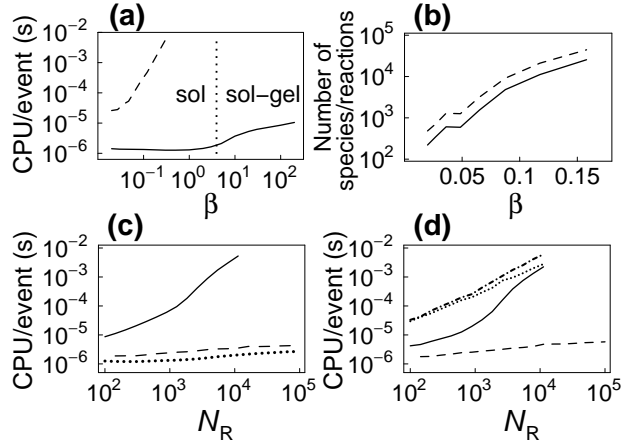}
\caption{\label{fig:efficiency} Efficiency of simulation of the TLBR model. 
(a) Dependence of CPU time per reaction event for rule-based KMC simulation (solid line) vs. on-the-fly simulation \cite{Fa05a,Lo05,Gi07} (dashed line).
(b) Effective network size as a function of $\beta$.  The solid and dashed lines indicate the numbers of species populated and reactions fired, respectively, in on-the-fly simulation. Calculations were performed using BioNetGen \cite{Bl04,Fa08}.  
(c) Dependence of CPU time per reaction event on $N_R$ for $\beta=50$ (solid line), $\beta=0.1$ (dashed line), and $\beta=50$ without connectivity checks (dotted line).  For $\beta=50$, the fraction of KMC steps that result in null events is approximately 0.6 for any value of $N_R$.  The fraction is essentially 0.0 for $\beta=0.1$.  Note that the system is above (below) the percolation transition at $\beta=50$ ($\beta=0.1$).
(d) Importance of null events.  The solid and dashed lines are calculated using auxiliary non-local component state information to minimize the cost of null events for $\beta=50$ and $\beta=0.1$, respectively.  The line broken in a dash-dot pattern and the dotted line are calculated using a problem-specific rejection-free procedure for $\beta=50$ and $\beta=0.1$, respectively.
Additional simulation parameters: 
(a) and (b) $N_R=300$, $N_L=4,200$, and $c=0.84$;  
(c) and (d) $N_L=14\, N_R$ and $c=0.84$.
The value of $k_{\text{off}}$ was held fixed at 0.01 s${}^{-1}$ in all simulations.  All reported results are based on simulation for 3,000 s after equilibration. 
}
\end{figure}

On-the-fly simulation is not adequate for simulating TLBR kinetics for many combinations of parameter values, especially for parameter values that favor the formation of large aggregates.  As shown in Fig.~\ref{fig:efficiency}(a), the cost of on-the-fly simulation becomes overwhelming at $\beta$ values far below the percolation transition because the number of species and reactions sampled during a simulation grows steeply with $\beta$ (Fig.~\ref{fig:efficiency}(b)). In contrast, the cost per reaction event of rule-based KMC is constant nearly up to the critical value of $\beta$.  Above the percolation transition, there is an increase in cost per reaction event that coincides with the growth in the average size of the largest aggregate, which depends on the number of molecules in the system.  As shown in Fig.~\ref{fig:efficiency}(c), there is a linear increase in the cost per reaction event with system size (as measured by number of receptors) above the percolation transition.  This increase can be attributed to the cost of enforcing the prohibition against cyclic aggregates, which requires checking the connectivity of two reacting sites, because when connectivity checks are omitted, the cost per reaction event remains constant in the sol-gel region (cf. solid and dotted lines in Fig.~\ref{fig:efficiency}(c)).  Connectivity checks are performed by breadth-first traversals of graphs representing ligand-receptor aggregates, which depend linearly on the number of vertices visited \cite{Th01}. 

%%%%%FIGURE 3

To investigate the effect of null events on simulation efficiency, we modified the simulation procedure to minimize the cost of null events. Null events arise from the step of evaluating the application condition of a rule.  The purpose of this step, in general, is to determine if components selected to potentially undergo a reaction on the basis of their local properties possess the non-local properties required of true reactants.  For rule $R_2$ of the TLBR model, the non-local property that reactants must possess is a lack of connectivity: two components are not allowed to bond if they are part of the same molecular complex.  By appending information about component membership in molecular complexes to local component states, we can use this non-local state information to determine connectivity when evaluating the application condition of $R_2$.  The frequency of null events is unchanged with this approach, which requires more programming effort, but null events associated with $R_2$ are less expensive. As shown in Fig.~\ref{fig:efficiency}(d), use of auxiliary information about component membership in complexes can speed simulation by 2- to 3-fold under conditions when large aggregates form, but scaling with system size is similar to the case when the auxiliary information is not used.  The linear increase in cost with system size occurs because graph traversal is required to update information about component membership in complexes whenever a ligand and receptor dissociate.  These results suggest that linear scaling with system size above the percolation transition is unavoidable and that the inherent features of the TLBR model play a more important role in determining the efficiency with which this model can be simulated than the incorporation of null events in the simulation procedure.

To further investigate the effect of null events on simulation efficiency, we implemented a problem-specific rejection-free method of simulation.  (The source code is available upon request.)  In this method, we essentially form the direct product of the sets $X_{21}$ and $X_{22}$, $X_2=X_{21} \times X_{22}$, and eliminate the set of non-reactive pairs of components, $\bar{X}_2$, from $X_2$, such that $r_2$ can be calculated as $(k_{+2}/V)|X_2 \backslash \bar{X}_2|$.  As illustrated in Fig.~\ref{fig:efficiency}(d), the cost of this approach scales linearly with system size both above and below the percolation transition, because the cost of finding a reactive pair of sites is proportional to the number of potentially reactive sites. In contrast, for the general-purpose procedure incorporating null events, cost is constant below the percolation transition and scales linearly with system size only above the percolation transition (Figs.~\ref{fig:efficiency}(c) and \ref{fig:efficiency}(d)).  These results suggest that null-event sampling provides both a simple and efficient means to evaluate and apply reaction rules that specify global features of reactants.

Our interest in developing a method to simulate models such as the TLBR model was prompted in part by the study of Posner et al. \cite{Po02}, who showed that a synthetic antigen with three symmetrically arrayed hapten groups generates a strong cellular secretory response through interaction with bivalent IgE antibody attached to cell-surface Fc$\epsilon$RI (the high-affinity IgE receptor), whereas the bivalent analogue of this antigen generates no secretory response.  Further motivation was provided by earlier studies indicating that the size of ligand-induced receptor aggregates as well as the kinetics of ligand-receptor binding are important factors that influence Fc$\epsilon$RI-mediated cellular responses to antigen \cite{Me92,Me02}.  The molecular mechanisms responsible for these effects, which are largely uncharacterized, may perhaps be identified with the help of models that capture the dynamics of ligand-induced receptor aggregation and receptor-mediated signaling events \cite{Go02,Fa03,Go04,Kh06}.  Analyses of such models require suitable simulation methods, which have not been available.

\begin{figure}
\includegraphics[width=2.2in]{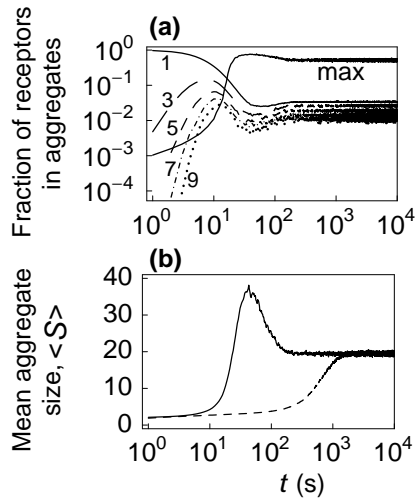}
\caption{\label{fig:results} Kinetics of the TLBR model. 
(a) Fraction of receptors in aggregates with 1, 3, 5, 7, or 9 receptors or in the largest aggregate as a function of time in the sol-gel coexistence phase ($N_L=50,000$ and $c=2.7$).
(b) Mean aggregate size as a function of time for same conditions as (a) (solid line) and at a lower ligand concentration (dashed line) that gives the same mean size at equilibrium ($N_L=2,000$ and $c=0.11$).
Additional simulation parameters: 
$N_R=3,000$, $\beta=16.8$, and $k_{\text{off}}=0.01$ s$^{-1}$. Results are averaged over 40 simulation runs.
Mean aggregate size is determined by 
$\left\langle S \right\rangle  = {\sum\nolimits_{i = 2}^{N_{R} } {i\,n_i } }/{\sum\nolimits_{i = 2}^{N_{R} } {n_i } }$, where $n_i$ is the number of aggregates containing $i$ receptors.  Parameter values were chosen arbitrarily for the purpose of demonstrating the rule-based KMC method, but they are expected to be somewhat reasonable for the case of a population of ligands, each with three 2,4-dinitrophenol (DNP) hapten groups, interacting with a population of monoclonal cell-surface anti-DNP IgE antibodies, each with two antigen-combining sites \cite{Po02,Po07}.
}
\end{figure}

Simulation of the aggregation kinetics of the TLBR model generates two predictions that could be relevant for understanding Fc$\epsilon$RI-mediated signaling, and cellular regulation in general, and that can be tested using available reagents \cite{Po02,Bi07,Po07,Si07}.  First, as seen in Fig.~\ref{fig:results}(a), small receptor aggregates may form transiently before the formation of a giant aggregate in the sol-gel region.  This result may have biological significance because small aggregates of Fc$\epsilon$RI (e.g., dimers and trimers) stimulate cellular responses \cite{Se77,Fe80}, whereas large aggregates of Fc$\epsilon$RI can be inhibitory \cite{Be73}.   Second, as seen in Fig.~\ref{fig:results}(b), two ligand doses that stimulate receptor aggregation to the same extent at equilibrium can generate qualitatively distinct time courses of receptor aggregation, which may have functional consequences.   For example, the two doses might elicit different early cellular responses but similar late cellular responses to the presence of ligand.  In any case, a characterization of the different signaling events triggered by the two doses could yield insights into temporal aspects of cellular signal processing.

The time courses of Fig.~\ref{fig:results}(b) are qualitatively different for the following reason.  For the parameters used in simulations, ligand capture is the rate-limiting step in ligand-induced receptor aggregation (i.e., ligand capture is slower than receptor cross-linking).  Furthermore, for the case of the higher ligand dose, the amount of bound ligand passes through an optimal level for receptor cross-linking during the transient.  When the kinetics of ligand capture are accelerated without changing equilibrium, the overshoot seen in Fig.~\ref{fig:results}(b) disappears (not shown).  One can be convinced that receptor aggregation is maximal at an optimal ligand concentration by considering the extremes of ligand and receptor excess.  When receptors are in large excess, ligands bind few receptors, and as a result, there is little cross-linking, even though each bound ligand tends to cross-link as many receptors as possible.  When ligands are in large excess, many receptors are bound, but each receptor tends to be bound to only a single ligand, because the pool of free ligand outcompetes the pool of bound ligand for free receptor sites.  The dependence of receptor aggregation on ligand concentration has been thoroughly studied by Goldstein and Perelson \cite{Go84}.  The results of this study can be used to select different ligand doses that yield the same level of receptor aggregation at equilibrium; the simulation method presented here can be used to reveal the dose-dependent kinetics (Fig.~\ref{fig:results}(b)).

%%%%%FIGURE 4

%%%% CLOSING REMARKS

\begin{figure}
\centering
\includegraphics[width=2.5in]{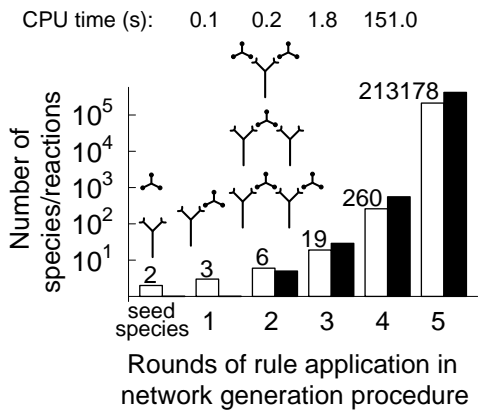}
\caption{\label{fig:GF} Generation of the reaction network implied by the rules of Fig.~\ref{fig:TLBR}. Starting from two speed species (free ligand and free receptor), successive rounds of rule application generate new chemical species and reactions.  In the process of network generation, species are represented by graphs and rule application is comprised of graph rewriting operations \cite{Bl06}. The two seed species and the four species generated in the first two rounds of rule application are illustrated using the conventions of Fig.~\ref{fig:TLBR}.  White bars indicate the number of species in the partially generated network at each step in the process of network generation. Black bars indicate the number of reactions. Indicated at top is the total CPU time required to perform each of the first four rounds of rule application using BioNetGen \cite{Bl04,Fa08} running on a desktop workstation.  CPU time is not reported for the fifth round of rule application, which was performed over the course of several days.}
\end{figure}

Large-scale reaction networks derived from rules strain the capabilities of conventional simulation methods \cite{Hl06}, which has hindered applications of the rule-based modeling approach and motivated efforts to make simulations of rule-based models more manageable, for example, by finding model reductions \cite{Bo05,Bo06,Co06,Ko07,Bo08}.  Indeed, even generating a reaction network from a set of rules can be an impractical process (Fig.~\ref{fig:GF}).  As indicated in Fig.~\ref{fig:GF}, the partial network generated from the rules of the TLBR model (Fig.~\ref{fig:TLBR}) after just five rounds of rule application consists of hundreds of thousands of chemical species and reactions.  However, this partial network is far from being large enough to account for the aggregates considered in Fig.~\ref{fig:results}.  The largest aggregate considered in the partial network contains just 16 receptors, whereas aggregates considered in Fig.~\ref{fig:results} contain about 20 receptors on average at equilibrium, with larger aggregates forming during the transient for the case of higher ligand concentration. 

%%%%%FIGURE 5

We have presented a method for simulating the kinetics of reaction rules that implicitly define a large-scale reaction network.  Development of this method was inspired by {\sc StochSim} \cite{Mo98,Sh01,Le01}, an early rule-based modeling software tool that implements a particle-based stochastic simulation method that has a cost independent of the number of reactions implied by rules. However, this method relies on an inefficient event sampling algorithm that produces a high fraction of unsuccessful moves (null events) for stiff systems.  A further drawback of the {\sc StochSim} framework, which prevents {\sc StochSim} from being used to simulate the TLBR model, is a limited ability to represent the connectivity of molecular complexes and to process rules that change molecular connectivity \cite{Hl06}.  The method presented here can be applied to simulate more expressive rules, and it takes advantage of the more efficient event sampling afforded by continuous time Monte Carlo methods \cite{Bo75}.  The method avoids null events arising from differences in the time scales of reactions (stiffness), but uses sampling with the introduction of null events to avoid forming the direct products of sets of potentially reactive components, which would incur a linear cost per reaction event with respect to system size for bimolecular reactions (Fig.~\ref{fig:efficiency}(d)).  For simulation of the TLBR model, below the percolation transition or without the connectivity condition of $R_2$, nearly constant scaling with system size is achieved (Fig.~\ref{fig:efficiency}(c)).  Above the percolation transition, linear scaling is observed because of the cost of enforcing the connectivity condition.

The challenges of simulating the TLBR model arise from the number of topologically distinct molecular complexes that become possible, and indeed populated, as average receptor aggregate size grows (Fig.~\ref{fig:efficiency}(b)).  In our experience, this type of problem commonly arises when attempting to model cellular regulatory systems, and we have shown for the first time how such problems related to aggregation can be solved.  It should be noted that the DFFK method has also been used to simulate the TLBR model as a test problem but without consideration of the connectivity condition of $R_2$ (W. Fontana, personal communication).  To properly consider cell-surface interactions between ligand and receptor, one must distinguish between intra- and intermolecular binding, which is enabled by the novel step in the procedure reported here that involves checking a rule application condition.  It should also be noted that related methods, involving assumptions similar to those typically made in a rule-based modeling approach, have recently been used to model epitaxial growth \cite{Sch04,Sch08}, self assembly \cite{Ja05,Zh06,Sw08}, and complex polymerization kinetics \cite{Ch07}, and thus, the approach described here is relevant for studying these types of physical systems as well as cellular regulatory systems. Rule-based KMC should be a useful tool for simulating a wide range of physical systems marked by combinatorial complexity, i.e., large reaction network size resulting from combinations of a relatively small number of molecular interactions. 

A potential application area of the rule-based KMC method is colloidal ferrofluids that undergo a self-assembly process and can form polymer-like linear chains or isotropic aggregates \cite{Tl00}. Another is associating polymers that play an important role in biological tissues \cite{Hi98}. These polymers form thermoreversible gels containing disordered supramolecular aggregates \cite{Ku99}.  Finally, we note that various complex phase behaviors have been explained with the help of thermodynamic models \cite{Ku99,Lu78,Zi03}. The rule-based KMC method could perhaps be used to extend these results and study the dynamics of the phase transitions in these systems.

\begin{acknowledgments}
We thank M. Challacombe, W. Fontana, I. Nemenman, M.E. Wall, and A. Zilman for reading the manuscript and providing constructive feedback.  We thank J. Colvin, V. Danos, J. Krivine, and R. G. Posner for helpful discussions. This work was supported by NIH grants RR18754 and GM076570 and DOE contract DE-AC52-06NA25396. J.Y. and J.R.F. acknowledge additional institutional support.  
\end{acknowledgments}

\bibliography{rbkmc}

\end{document}